\pdfoutput=1
\documentclass[conference,9pt]{IEEEtran}
\usepackage{algpseudocode}                                 
\usepackage{algorithm}
\usepackage{graphicx}                                      
\usepackage{amsmath}
\usepackage{amssymb}
\usepackage{amsfonts}
\usepackage{amsthm}
\usepackage[mathcal]{eucal}
\usepackage{booktabs}
\usepackage{multirow}
\usepackage[subrefformat=parens,farskip=0pt,justification=centering]{subfig}
\usepackage{color}
\usepackage{cite}                                          
\usepackage{comment}                                       
\usepackage{soul}                                          
\soulregister\cite7
\soulregister\ref7
\soulregister\pageref7
\usepackage{etoolbox}                                      
\usepackage{url}
\usepackage{nth}                                           
\usepackage{bm}                                            
\usepackage{courier}
\usepackage{balance}
\usepackage{threeparttable}
\usepackage[bookmarks=false]{hyperref}
\hypersetup{
    colorlinks = true,
    citecolor  = blue,
    linkcolor  = blue,
    urlcolor   = blue,
}
\usepackage{tikz}
\usetikzlibrary{positioning, fit}
\usepackage{filecontents}                                  
\usepackage{pgfplots}
\usepackage{pgfplotstable}
\pgfplotsset{compat=newest}
\usepackage{caption}
\usepackage{cleveref}
\Crefformat{figure}{Fig.~#2#1#3}                           
\Crefname{subfigure}{Fig.}{Figs.}
\Crefname{figure}{Fig.}{Figs.}
\Crefformat{table}{TABLE~#2#1#3}                           
\definecolor{CUHKorange}{RGB}{244,106,18} 
\definecolor{CUHKblue}{RGB}{0,111,190}    
\definecolor{CUHKgreen}{RGB}{0,127,128}   
\definecolor{CUHKred}{RGB}{228,46,36}     
\definecolor{CUHKyellow}{RGB}{198,148,34} 
\definecolor{CUHKdark}{RGB}{114,44,114}   
\definecolor{CUHKmiddle}{RGB}{144,44,144} 
\definecolor{CUHKlight}{RGB}{167,44,167} 

\newcommand{\calN}{\mathcal{N}}

\newcommand{\calP}{\mathcal{P}}

\renewcommand{\vec}[1]{\boldsymbol{#1}}

\usepackage[top=0.75in,bottom=0.75in,left=0.55in,right=0.55in]{geometry}
\newcommand{\subparagraph}{}
\usepackage{titlesec}
\titlespacing*{\section}{0pt}{1.8ex plus .2ex minus .2ex}{0.4ex plus .2ex}
\titlespacing*{\subsection}{0pt}{1.0ex plus .2ex minus .2ex}{0.2ex plus .2ex}

\renewcommand{\vec}[1]{\boldsymbol{#1}}    
\algrenewcommand\textproc{\texttt}

\makeatletter
\let\OldStatex\Statex
\renewcommand{\Statex}[1][3]{%
  \setlength\@tempdima{\algorithmicindent}%
  \OldStatex\hskip\dimexpr#1\@tempdima\relax
}
\makeatother

\RequirePackage[normalem]{ulem} 
\RequirePackage{color}\definecolor{RED}{rgb}{1,0,0}\definecolor{BLUE}{rgb}{0,0,1} 

\graphicspath{{./figs/}}

\IEEEoverridecommandlockouts

\definecolor{mypurple}{RGB}{226,183,255}  
\definecolor{myblue}{RGB}{178,179,249}    
\definecolor{mydarkblue}{RGB}{0,107,250}  
\definecolor{mygrey}{RGB}{126,126,126}    
\definecolor{myred}{RGB}{228,46,36}       
\definecolor{myorange}{RGB}{244,106,18}   
\definecolor{myyellow}{RGB}{198,148,34}   
\definecolor{CUHKorange}{RGB}{244,106,18} 
\definecolor{CUHKblue}{RGB}{0,111,190}    
\definecolor{CUHKgreen}{RGB}{0,127,128}   
\definecolor{CUHKred}{RGB}{228,46,36}     
\definecolor{CUHKyellow}{RGB}{198,148,34} 
\definecolor{CUHKdark}{RGB}{114,44,114}   
\definecolor{CUHKmiddle}{RGB}{144,44,144} 
\definecolor{CUHKlight}{RGB}{167,44,167}  

\begin{document}

\title{
OpenMPL: An Open Source Layout Decomposer
\\ \Large (Invited Paper)
}
\author{
    \IEEEauthorblockN{
        Wei Li\IEEEauthorrefmark{1},
        Yuzhe Ma\IEEEauthorrefmark{1},
        Qi Sun\IEEEauthorrefmark{1},
        Yibo Lin\IEEEauthorrefmark{2},
        Iris Hui-Ru Jiang\IEEEauthorrefmark{3},
        Bei Yu\IEEEauthorrefmark{1},
        David Z.~Pan\IEEEauthorrefmark{4}
    }
    \IEEEauthorblockA{
        \IEEEauthorrefmark{1}The Chinese University of Hong Kong, \ \ \ \ 
        \IEEEauthorrefmark{2}Peking University \\ 
        \IEEEauthorrefmark{3}National Taiwan University, \ \ \ \ 
        \IEEEauthorrefmark{4}University of Texas at Austin\\
        Email: \normalsize\texttt{\{wli,byu\}@cse.cuhk.edu.hk, yibolin@pku.edu.cn}
    }
}

\maketitle
\thispagestyle{empty}

\begin{abstract}
    Multiple patterning lithography has been widely adopted in advanced technology nodes of VLSI manufacturing. 
    As a key step in the design flow, multiple patterning layout decomposition (MPLD) is critical to design closure. 
    Due to the $\calN\calP$-hardness of the general decomposition problem, various efficient algorithms have been proposed with high quality solutions. 
    However, with increasingly complicated design flow and peripheral processing steps, 
    developing a high-quality layout decomposer becomes more and more difficult, 
    slowing down the further advancement in this field. 
    This paper presents $\mathsf{OpenMPL}$ \cite{TOOL-OpenMPL}, an open-source layout decomposition framework, 
    with well-separated peripheral processing and the core solving steps. 
    We demonstrate the flexibility of the framework with efficient implementations of various state-of-the-art algorithms, which enable us to reproduce most of the recent results on widely-recognized benchmarks. 
    We believe $\mathsf{OpenMPL}$ can pave the road for developing layout decomposition engines and stimulate further researches on this problem. 
    
\end{abstract}

\section{Introduction}
\label{sec:intro}


Multiple patterning layout decomposition (MPLD) has been adopted to improve the lithography resolution.
The key idea of MPLD is to assign features that are close to each other to different masks, such that within each mask,
the features are far away enough to be printed with existing lithography techniques.
MPLD can be divided into double patterning layout decomposition (DPLD), triple patterning layout decomposition (TPLD) and quadruple patterning layout decomposition (QPLD) according to the number of masks.
This problem is difficult since it is a variation of the graph coloring problem, which is $\calN\calP$-hard for $k\geq 3$, where $k$ is the number of masks. 

\begin{figure}[tb!]
	\centering
    \subfloat[]{\includegraphics[width=.32\linewidth]{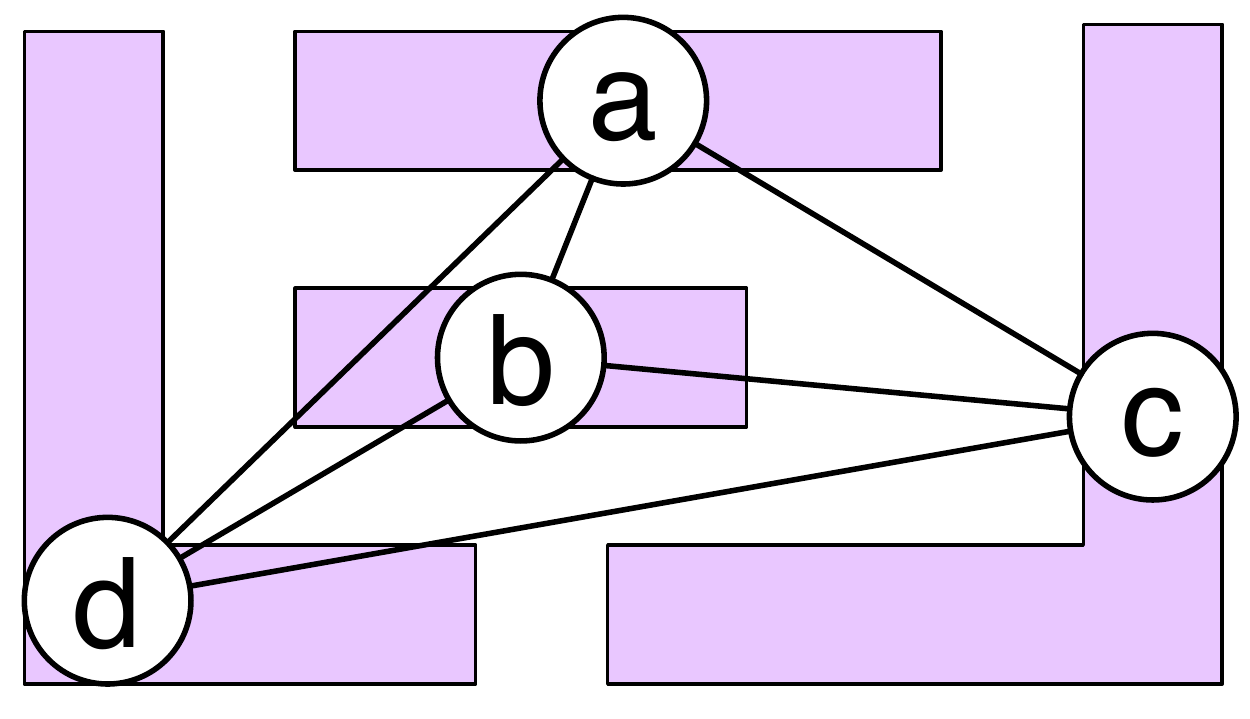}\label{fig:stitch1}} \hspace{.2in}
    \subfloat[]{\includegraphics[width=.32\linewidth]{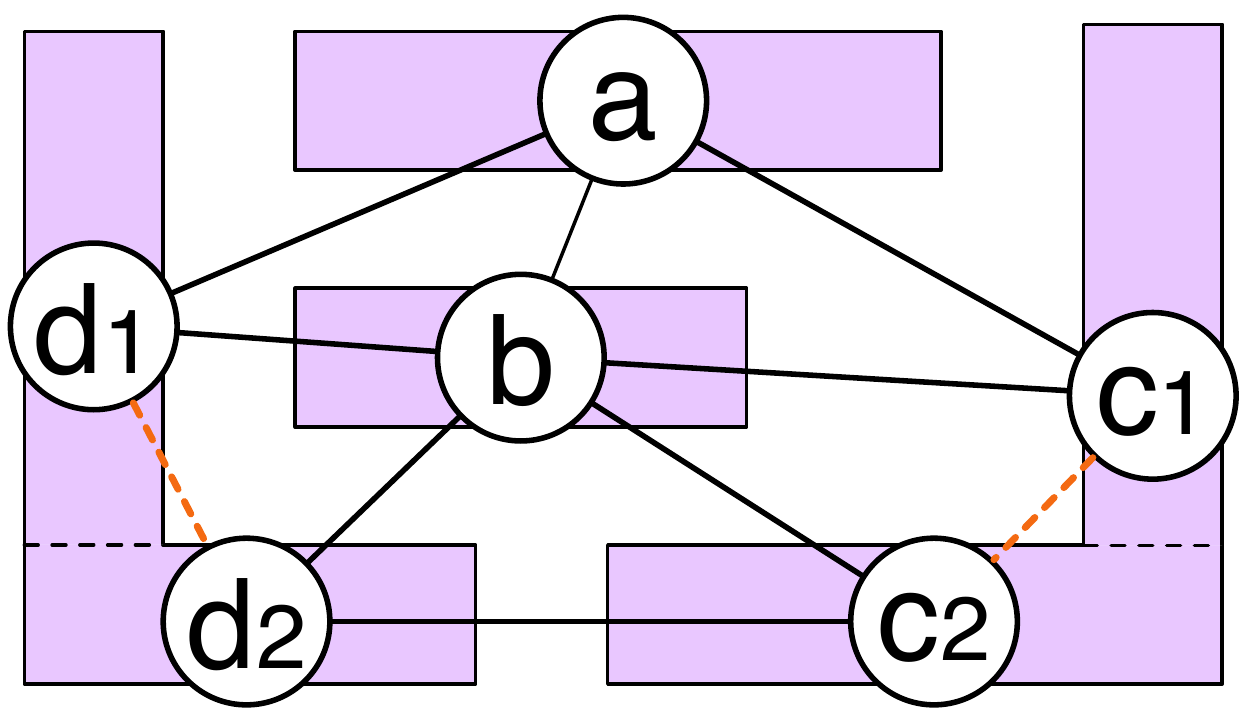}\label{fig:stitch2}}\\
    \subfloat[]{\includegraphics[width=.32\linewidth]{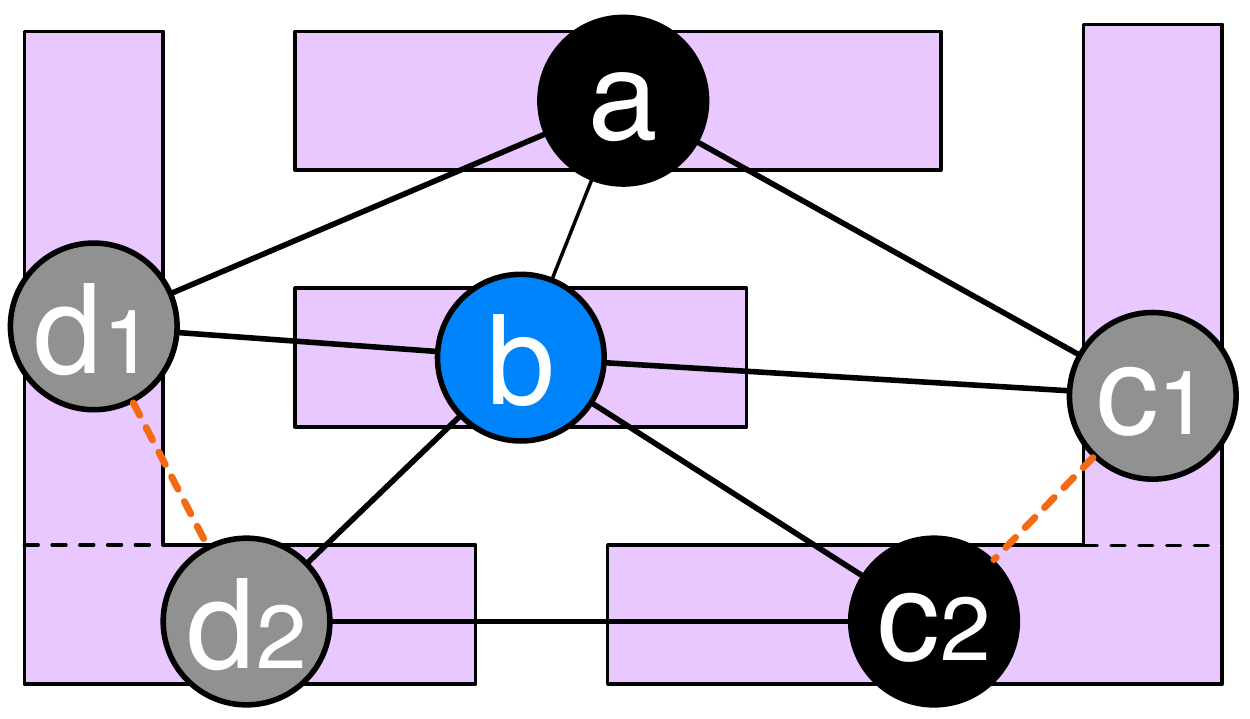}} \hspace{.2in}
    \subfloat[]{\includegraphics[width=.32\linewidth]{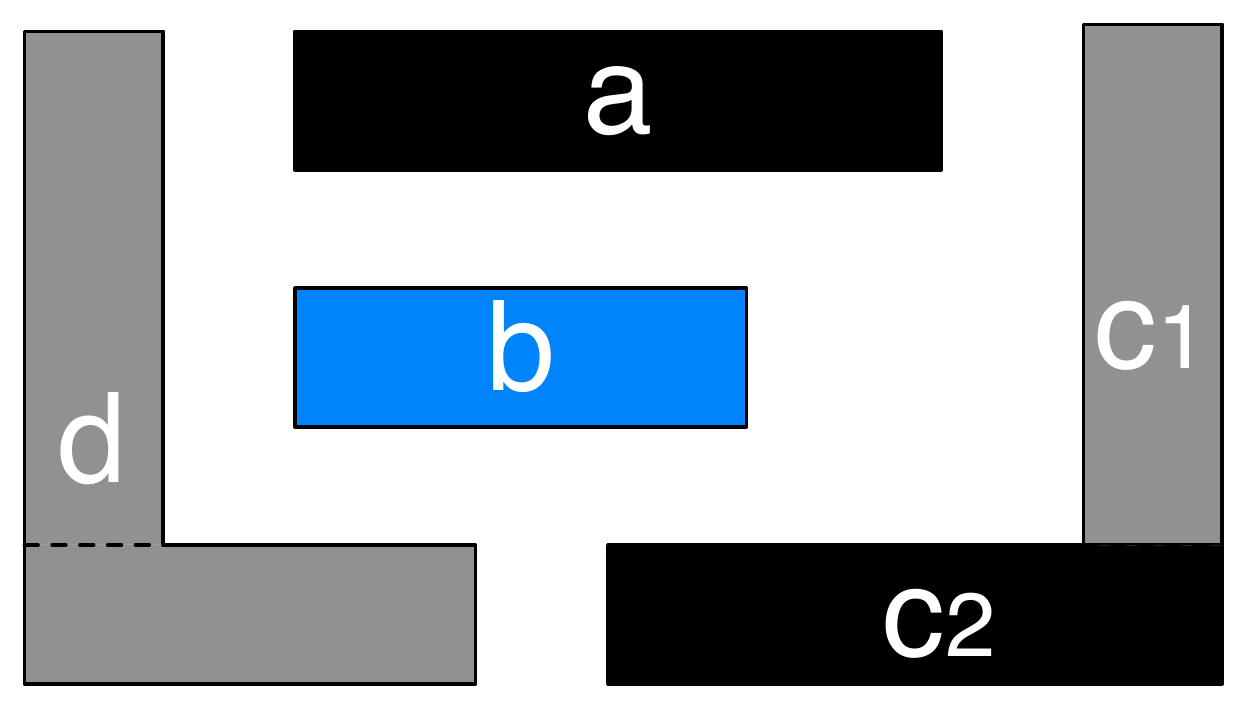}}
    \caption{ An example of TPLD with stitches.
		(a) The constructed layout graph without stitch candidates insertion, which is actually a 4-clique and therefore not 3-colorable;
		(b) The constructed layout graph with stitch candidates insertion. Two stitch candidates are introduced and the original 4-clique is dismissed;
		(c) Coloring on the layout graph with stitch candidates insertion. No conflict in the final coloring result with only one stitch inserted;
		(d) The final decomposed layout with three masks (each color corresponds to one mask), the cost is 0.1 following \Cref{formula:mpl}.
	}
	\label{fig:stitch}
\end{figure}

\Cref{fig:stitch} is an example of TPLD, where three colors represent corresponding masks.
Different from classical graph coloring problem, the layout decomposition problem has several unique characteristics. 
1) Stitch: a polygon feature is allowed to be split into multiple overlapping segments to resolve coloring conflicts, as shown in the dashed edge of \Cref{fig:stitch2}.
2) Special patterns: circuit layout follows some kinds of patterns due to the design styles, e.g., the alternative power and ground lines that may help to simplify the graph. 
3) Complex rules: besides the widely-adopted spacing constraint for the same color, there are also other rules like the different color spacing constraints \cite{MPL-USP2017-Chen} related to the ordering of masks, and the pre-colored constraints where the colors of some features are pre-determined before decomposition. 
These characteristics make the problem special and require customized algorithms to solve it effectively and efficiently. 

To achieve high efficiency and meanwhile maintain high solution quality, a variety of decomposition algorithms have been proposed.
These algorithms can be roughly categorized into three types: mathematical programming and relaxation, graph-theoretical approaches and search-based approaches \cite{MPL-DAC2015-Pan,MPL-VLSI-SOC2017-Ma}.
Mathematical programming is to solve MPLD problem by formulating it into a mathematical programming model, such as integer linear programming (ILP) for DPLD \cite{DPL-ICCAD2009-Xu,DPL-TCAD2010-Kahng,DPL-TCAD2010-Yuan} and TPLD \cite{TPL-TCAD2015-Yu,TPL-ICCAD2013-Yu,TPLEC-JM3-2015-Yu}. 
Due to the $\calN\calP$-hardness, relaxation techniques such as semidefinite programming (SDP) \cite{TPL-TCAD2015-Yu}, linear programming (LP) \cite{TPL-JM3-2017-Lin} and a discrete relaxation method \cite{TPL-TC2017-Li} are also proposed based on ILP. 
Another category is to directly perform color assignment based on a set of graph-theoretical algorithms,
e.g., the maximal independent set (MIS) \cite{TPL-TCAD2014-Fang}, the shortest-path \cite{TPL-ISPD2015-Chien,TPL-ICCAD2012-Tian}, and the fixed-parameter tractable (FPT) \cite{DPL-DAC2017-Kuang} algorithms. 
Search-based algorithms follow a divide-and-conquer principle with each sub-graph containing less than 20 nodes. 
Then a search procedure is applied to find the optimal solutions for small sub-graphs \cite{TPL-SPIE2014-Yu,TPL-DAC2013-Kuang,TPL-TCAD2014-Fang,TPL-ICCAD2013-Zhang,TPL-TCAD2015-Yu,TPL-DAC2016-Chang,TPL-SPIE2014-Fang}. 
Besides the researches on single layout decomposition stage, recent work \cite{OPC-ICCAD2017-Ma} pioneers a new direction which integrates layout decomposition and mask optimization seamlessly, achieving compelling results from a global view of the solution space. 


Besides the innovations to the core algorithms, many graph simplification techniques have been developed to reduce the problem size, 
such as independent component computation (ICC) \cite{TPL-TCAD2015-Yu}, 
hide small degrees \cite{TPL-DAC2013-Kuang,TPL-TCAD2015-Yu}, 
biconnected component analysis \cite{DPL-TCAD2010-Kahng,DPL-TCAD2010-Yuan}, 
sub-K4 structure merging for TPLD \cite{TPL-JM3-2017-Lin}. 

To reduce the repeated effort in reimplementation of the whole decomposition framework and lower the bar of the research on MPLD, in this paper, we present a general and open-source framework, $\mathsf{OpenMPL}$,
as an open platform for developing MPLD algorithms. We carefully design the software architectures and APIs to decouple the innovations on the core optimization steps. 
For example, one can focus on developing novel graph simplification or decomposition techniques without worrying about the peripheral processing issues as the platform provides clean and well-defined APIs for kernel optimization engines. 
We also provide efficient implementations of widely-adopted graph simplification techniques and state-of-the-art layout decomposition algorithms
which have been introduced above.
We believe that this open platform paves the road for the development of MPLD engines and will stimulate more researches in the near future, eventually contributing to better manufacturability in advanced technology nodes. 

\section{The $\mathsf{OpenMPL}$ Framework}
\label{sec:tech}

\begin{figure}[tb!]
    \centering
    \includegraphics[width=1\linewidth]{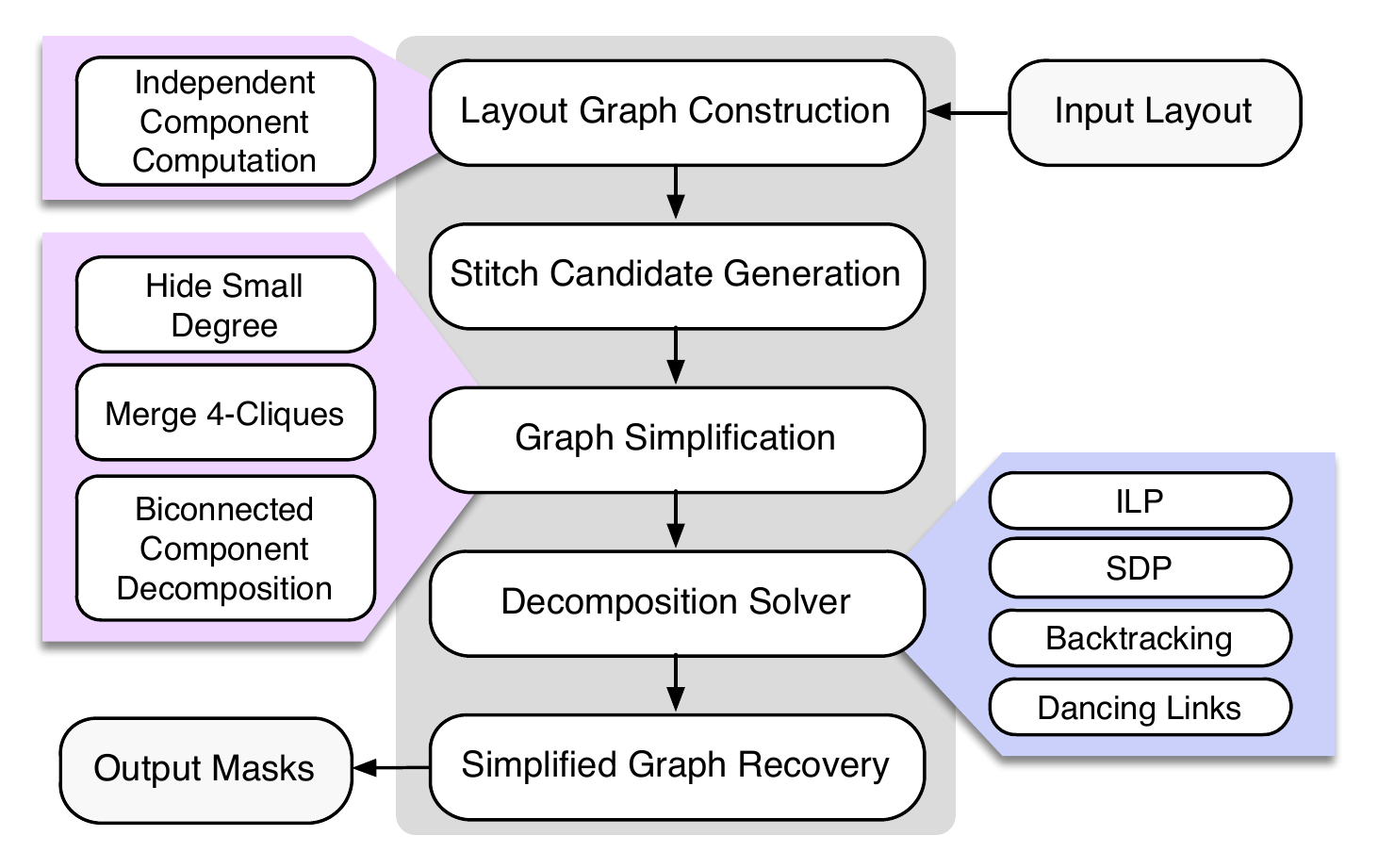}  
    \caption{The workflow of $\mathsf{OpenMPL}$.}
    \label{fig:flow}
\end{figure}

$\mathsf{OpenMPL}$ is a general framework offering various decomposition algorithms and different simplification techniques.
These methods are well embedded into the framework with unified interfaces.
The framework operation flow is depicted in \Cref{fig:flow}. Technical details are as follows.

\subsection{Problem Formulation}
Given an input layout specified by features in polygonal shapes, the layout can be translated into 
an undirected layout graph $G=(V,E)$, where every node $v_i \in V$ corresponds to one feature in layout and each edge $e_{ij} \in E$ is used to characterize the relationships between features. 
Considering conflict and stitch relationships, $E$ is composed of this two kinds of edges, denoted by 
$E=\{CE \cup SE\}$, where $SE$ is the set of stitch edges and $CE$ is the set of conflict edges.
The MPLD problem can be formulated as below: 
\begin{subequations}  \label{formula:mpl}
    \begin{align}
        \underset{\vec{x}}{\min} \ \ & \sum   c_{ij} + \alpha\sum s_{ij},   \\
        \textrm{s.t.} \ \ 
        & c_{ij} = (x_i==x_j),          && \forall e_{ij} \in CE, \\
        & s_{ij} = (x_i \neq x_j),      && \forall e_{ij} \in SE, \\
        & x_i \in \{0, 1, \ldots ,k\},  && \forall x_i \in \vec{x},
    \end{align}
\end{subequations}
where $x_{i}$ is a variable for the $k$ available colors of the pattern $v_{i}$, $c_{ij}$ is a binary variable representing conflict edge $e_{ij}\in CE$, 
$s_{ij}$ stands for stitch edge $e_{ij}\in SE$
, $\alpha$ is a user-defined parameter and is set as 0.1 by default in our framework to assign relative importance between the conflict cost and the stitch cost.
If two nodes, $x_{i}$ and $x_{j}$, within the minimal coloring distance are assigned the same color (i.e.~$x_i==x_j$), then $c_{ij}=1$.
On the contrary, $s_{ij}=1$ when two nodes connected by stitch edge are assigned different color (i.e.~$x_i \neq x_j$).
The objective function is to minimize weighted summation of the conflict number and the stitch number. 

\subsection{Design Principles}
$\mathsf{OpenMPL}$ is designed for end-users, developers and researchers as a general platform for MPLD algorithms. 
Therefore, we emphasize usability, efficiency, and extensibility during development. 
The core design principles are highlighted as follows. 
\begin{itemize}
    \item Decoupled design stages. 
        The implementation clearly separates different optimization stages in \Cref{fig:flow} 
        such that the interdependency between them is minimized. 
        In this way, developers can focus on verifying individual stages without worrying about cross-stage impacts. 
    \item Graphs for communications among kernel stages. 
        After layout graph construction, the graph simplification, decomposition solver, and the simplified graph recovery stages use pure graphs as input/output, without involving mask data. 
        This design leads to well-defined and highly separatable core algorithms, making the framework highly extensible. 
    \item Efficiency and generality for different mask data. 
        As a mask layer can be a contact layer or a metal layer, the processing efficiency varies significantly for different types of layers. 
        We design a general mask database with separate processing routines for contact layers and metal polygon layers for efficiency enabled by C++ polymorphism, 
        since contact layers can be processed in a much simpler way.
\end{itemize}

\subsection{Workflow and Functionalities}
The workflow of $\mathsf{OpenMPL}$ is shown in \Cref{fig:flow}. 
Firstly, one chip layout information (GDS) file is loaded and transformed into a layout graph, which is represented by a vector of rectangle pointers, where the rectangles are defined in $\mathsf{Boost}$. 
Secondly, a stitch insertion process \cite{DFMP-TCAD2015-Yu} is executed to generate a decomposed graph with stitches after simple ICC simplification.
Then the decomposed graph is simplified by several simplification techniques, where some of them are implemented in third-party library $\mathsf{Limbo}$ \cite{TOOL-Limbo}. 
After simplification, a coloring solver is called for each component in the decomposed graph to solve the component coloring problem.
Finally, our framework recovers nodes removed in the simplification step and assigns legal color for each removed node.
In the following sub-sections, we are going to introduce all of the functionalities in two crucial procedures of $\mathsf{OpenMPL}$, layout simplification and decomposition.

Layout graph simplification techniques can be used to reduce the graph size and therefore reduce the computational complexity.
We only need to deal with the smaller graph without affecting the final result.  
All of the four simplification techniques mentioned in \Cref{sec:intro} are supported in our framework, including \texttt{ICC}, \texttt{HIDE\_SMALL\_DEGREE}, \texttt{BICONNECTED\_COMPONENT} and \texttt{MERGE\_SUBK4}.


\texttt{ICC} is proposed based on the fact that there are many isolated clusters in a real layout such that \texttt{ICC} can break down the layout graph into several independent components. 
\texttt{HIDE\_SMALL\_DEGREE} temporarily removes the nodes whose degree is less than the number of color options in an iterative manner.
\texttt{BICONNECTED\_COMPONENT} simplifies the layout graph by removing all the bridge vertices. 
\texttt{MERGE\_SUBK4} detects and merges specific structures whose number of edges is exactly one less than four-clique structures and thus is only applicable for TPLD.
Different simplification techniques require different recovery methods.
However, those nodes which are shared among different components may be assigned different colors after recovery.
To tackle this, color rotation is implemented in our framework.


Graph color assignment is the most crucial step in the flow, which impacts the final coloring results directly.
In graph color assignment, simplified graph is provided and each vertex in the graph should be assigned one color by specified algorithm. 
Users of our framework can specify which algorithm is adopted and the number of colors available.
$\mathsf{OpenMPL}$ has supported all of the commonly-used algorithms in layout decomposition. 
The methodologies are briefly introduced in the following context.
\begin{itemize}
    \item \textbf{Integer Linear Programming}:
        Solving Problem \eqref{formula:mpl} is non-trivial. 
        A widely used method to solve this problem is integer linear programming \cite{DPL-ICCAD2009-Xu,DPL-TCAD2010-Kahng,DPL-TCAD2010-Yuan,TPL-TCAD2015-Yu}, 
        which converts this problem into linear programming by binary encoding of vertex colors and replacing the color constraints with a set of inequality constraints.
        ILP can be easily extended to solve different coloring problems, including TPLD and QPLD.
        Our framework is based on the theory proposed in \cite{TPL-TCAD2015-Yu}.
        We use $\mathsf{Gurobi}$ \cite{TOOL-gurobi}, $\mathsf{Lemon}$ \cite{TOOL-lemon} and $\mathsf{CBC}$ \cite{TOOL-cbc} as the ILP solvers.
    

    \item \textbf{Semidefinite Programming}: 
        The discrete integer programming solving process of \Cref{formula:mpl} is $\calN\calP$-hard, thus it may suffer from run-time overhead for practical designs.
        As shown in \cite{TPL-TCAD2015-Yu,TPL-ISAAC2014-Tomomi,TPL-DAC2014-Yu}, the color assignment can be formulated as a vector programming and then relaxed and 
        solved by semi definite programming, which can be solved in polynomial time. 
        Given the solutions of SDP, a mapping process is used to map the solutions to coloring results.
        $\mathsf{CSDP}$ \cite{TOOL-csdp} is used as the SDP solver.
    
    \item \textbf{Backtracking}: 
    Backtracking is a DFS fashion algorithm that is used to find solutions with constraints in the whole solution space. 
    We also provide backtracking routine in $\mathsf{OpenMPL}$. 
    Though backtracking is widely used, its runtime performance is unsatisfactory for complicated graphs. 
    Therefore we use it as a sub-solver to solve color assignment problem in simple sub-graphs . 
    

    \item \textbf{Dancing Links}: 
    Different from original dancing links and algorithm X, dancing links based algorithm for MPLD problem \cite{TPL-DAC2016-Chang} concludes conflicts earlier by traversal in BFS manner, treating this problem as an exact cover problem. 
    In $\mathsf{OpenMPL}$, this solver is developed by ourselves instead of third-party libraries.
    We follow the statement in \cite{TPL-DAC2016-Chang} and build a classical dancing links data structure in our implementation so that there are still much room for optimization.
\end{itemize}
$\mathsf{OpenMPL}$ also supports decomposition algorithms like MIS, LP, etc, which cannot solve the graph containing stitch edges while work well on stitch-free graphs. Due to page limit, we leave the details on the tool release page \cite{TOOL-OpenMPL}. 

\subsection{Additional Features}

Some additional features are also supported for better usability, efficiency and extensibility.
    (1) $\mathsf{OpenMPL}$ supports \textbf{multi-threading} operation by $\mathsf{OpenMP}$ \cite{TOOL-OpenMP} and users can specify the number of threads. 
    Graph components are solved in parallel and layout decomposition algorithms also support multi-threading computations.
    (2) We can identify all the possible positions of stitches through pattern projections \cite{TPL-TCAD2015-Yu} in \textbf{stitch insertion}, 
    which is one of the most critical steps to parse a layout.
    One example of stitch is shown in \Cref{fig:stitch}. 
    There are lots of candidate positions for stitch insertion, and we only choose some stitches from those candidates. 
    One thing should be noted is that all the coloring algorithms provided in the framework share the identical stitch candidate generation procedure, which results in identical graph simplification results regardless of coloring algorithms. 
    (3) In practice, a pattern in the layout may be a polygon or rectangle.
    Consequently, the storage may vary from case to case.
    $\mathsf{OpenMPL}$ provides a \textbf{shape-friendly} system considering this case and users can 
    specify the shape, \texttt{POLYGON} or \texttt{RECTANGLE}, to guarantee the performance to avoid unnecessary calculations. 
    For polygonal inputs, to simplify the storage structure design and save space, $\mathsf{OpenMPL}$ firstly decomposes the polygons to rectangles.
    After reading the whole input file, DFS is utilized to find connected components and re-union rectangles into polygons.
    For rectangle circuits, we directly store these patterns without further operations.

\section{Experimental Results}
\label{sec:result}





\begin{table*}[!tb]
	\centering
	\caption{Decomposition Cost Comparison}
	\label{tab:total_result}
    \resizebox{17.6cm}{!} {
        \begin{tabular}{c|cccc|cccc|cccc|cccc}
            \toprule
            Circuit       &\multicolumn{4}{c|}{ILP}&\multicolumn{4}{c|}{SDP}&\multicolumn{4}{c|}{Dancing links}&\multicolumn{4}{c}{Backtracking}\\
            \cline{2-17}    &time(s)   &st\# &cn\# &cost  &time(s)&st\# &cn\# &cost&time(s)&st\# &cn\# &cost&time(s)&st\# &cn\# &cost\\ 
            \hline
            \texttt{C432}&  0.045 &4&0&0.4&  0.021 &4&0&0.4&  0.008 &4&0&0.4&  0.023 &4&0&0.4\\
            \texttt{C499}&  0.047 &0&0&0&  0.016 &0&0&0&  0.014 &0&0&0&  0.012 &0&0&0\\
            \texttt{C880}&  0.053 &7&0&0.7&  0.018 &7&0&0.7&  0.008 &7&0&0.7&  0.010 &7&0&0.7\\
            \texttt{C1355}&  0.050 &3&0&0.3&  0.022 &2&1&1.2&  0.013 &3&0&0.3&  0.035 &3&0&0.3\\
            \texttt{C1908}&  0.039 &1&0&0.1&  0.010 &1&0&0.1&  0.008 &1&0&0.1&  0.018 &1&0&0.1\\
            \texttt{C2670}&  0.064 &6&0&0.6&  0.025 &6&0&0.6&  0.016 &6&0&0.6&  0.024 &6&0&0.6\\
            \texttt{C3540}&  0.067 &8&1&1.8&  0.035 &8&1&1.8&  0.016 &8&1&1.8&  1.981 &8&1&1.8\\
            \texttt{C5315}&  0.074 &9&0&0.9&  0.029 &9&0&0.9&  0.012 &10&0&1&  0.034 &9&0&0.9\\
            \texttt{C6288}&  0.679 &205&1&21.5&  0.137 &200&6&26&  0.082 &203&5&25.3&  125.562 &205&1&21.5\\
            \texttt{C7552}&  0.105 &21&1&3.1&  0.049 &20&2&4&  0.040 &22&1&3.2&  0.143 &21&1&3.1\\
            \texttt{S1488}&  0.184 &2&0&0.2&  0.020 &2&0&0.2&  0.005 &2&0&0.2&  0.013&2&0&0.2\\
            \texttt{S38417}&  0.848 &55&19&24.5&  0.327 &50&24&29&  0.248 &56&20&25.6&  2.388 &55&19&24.5\\
            \texttt{S35932}&  2.756 &40&44&48&  0.910 &24&60&62.4&  0.513 &45&46&50.5&$>$3600&NA&NA&NA\\
            \texttt{S38584}&  2.183 &111&42&53.1&  0.908 &99&54&63.9&  0.475 &110&44&55&$>$3600&NA&NA&NA\\
            \texttt{S15850}&  2.148 &98&34&43.8&  0.852 &90&42&51&  0.470 &102&35&45.2&2340.87&98&34&43.8\\
            \hline
            average           &0.623 &38.000 &9.467 &13.267    &0.225  &34.800 &12.667 &16.147 &0.129 &38.600 &10.133&13.993&$>$647.32&NA &NA &NA\\
            ratio             &{1.000} &{1.000} &{1.000} &1.000&{0.361} &{0.916} &{1.338} &1.217&{0.206} &{1.016} &{1.070}&1.054&$>$1039.04&{NA} &{NA} &{NA}\\
            \bottomrule
        \end{tabular}
    }
\end{table*}


We implement $\mathsf{OpenMPL}$ in C++ and use $\mathsf{Boost}$ \cite{TOOL-boost} as the basic graphics library. 
All of the experiments are tested on an Intel Core 2.9 GHz Linux machine. 
We adapt scaled down and modified ISCAS benchmarks from \cite{TPL-TCAD2015-Yu} to conduct experiments, which are widely used in previous works. 
The minimum coloring spacing is set to 120$nm$ for the first ten cases and as 100$nm$ for the last five
cases, as in \cite{TPL-TCAD2014-Fang,TPL-TCAD2015-Yu,TPL-DAC2016-Chang}. The thread number is 8 and the graph simplification level is 3 (\texttt{ICC}, \texttt{HIDE\_SMALL\_DEGREE}, \texttt{BICONNECTED\_COMPONENT}). 
Fig. 16 shows the decomposition result for case \texttt{C432} by dancing links based algorithm, which can be obtained in 0.008 seconds.

\begin{figure}[!tb]
  \centering
  \includegraphics[width=.86\linewidth]{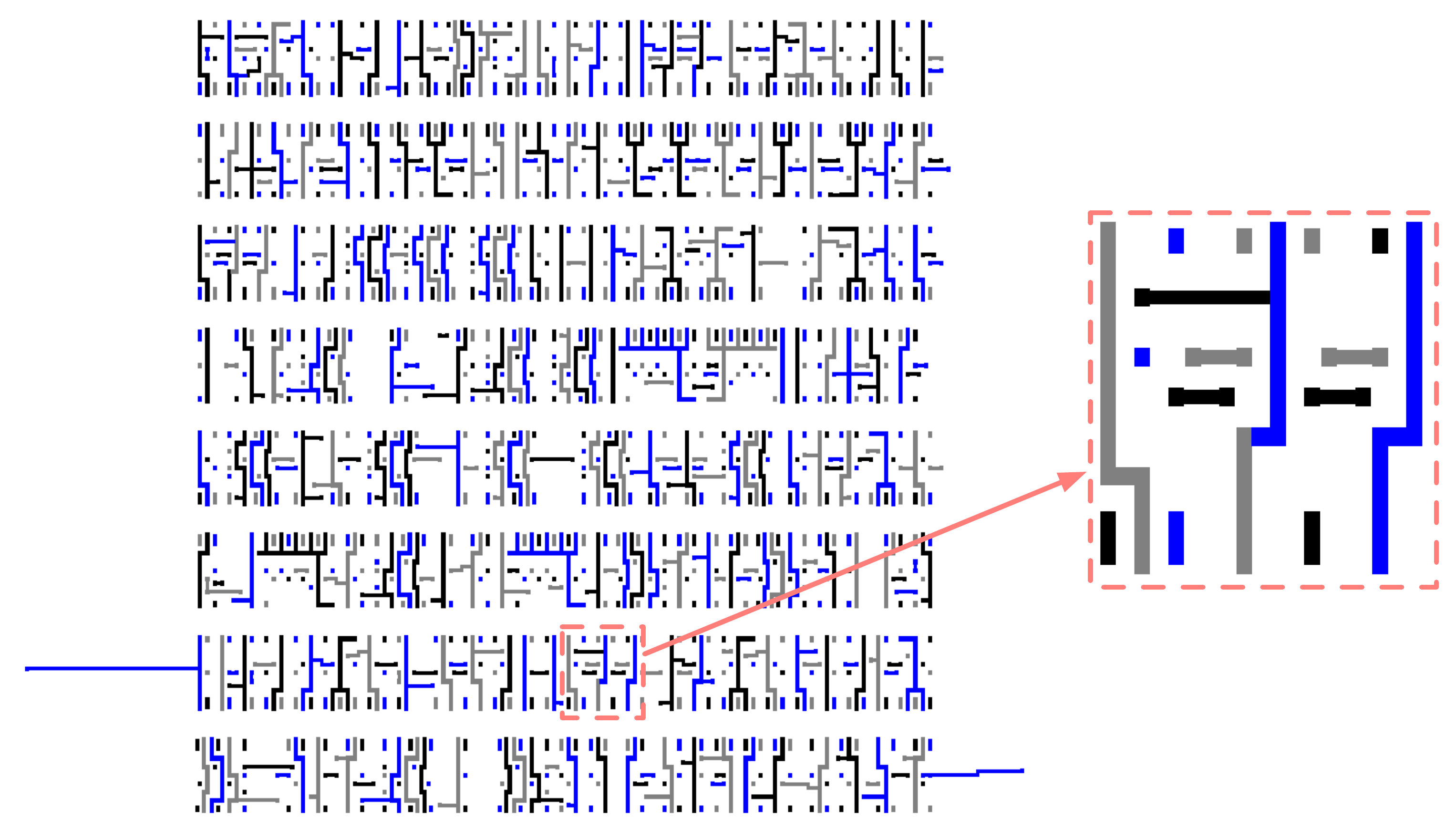}  
  \caption{Case \texttt{C432} decomposition result.}
  \label{fig:result-1}
\end{figure}

In this section, we only focus on the results of different decomposition algorithms with stitch insertion on TPLD problem due to page limit, 
which is more difficult to obtain optimal results compared with DPLD and stitch-free problems.
More detailed results and discussions can be found in \cite{TOOL-OpenMPL}.

Firstly, we discuss the effectiveness of different decomposition algorithms implemented in our framework.
Specially, MIS and LP are not listed since they cannot solve the problem with stitch edges. 
We compare the conflict and stitch number in coloring results and compute the corresponding cost following \Cref{formula:mpl}, where
parameter $\alpha$ is set to 0.1, thus the decomposition cost is calculated by $\text{cn}\# + 0.1 \cdot \text{st}\#$.

\Cref{tab:total_result} compares the performance of implemented algorithms, where ``ILP'', ``SDP'', ``Dancing links'' and ``Backtracking''
represent $\mathsf{OpenMPL}$ with corresponding algorithms respectively.
Columns ``st\#'' and ``cn\#'' denote the stitch number and the conflict number in the final coloring result, while the column ``cost'' is the decomposition cost. 

For the cases whose runtime are more than 3600 seconds, we directly terminate the computations and do not measure the effectiveness.
According to the results shown in the table, ILP-based algorithm achieves the best results no matter in the conflict number or the decomposition cost.
Also, the backtracking algorithm can achieve same results with ILP-based algorithm in the cases which can be solved within 3600 seconds by backtracking algorithm.
This is because all of the algorithms use the same simplification graphs as input and both ILP-based algorithm and backtracking can search for the optimal solution of 
\Cref{formula:mpl} in the search space.
The result of SDP-based algorithm is close to the optimal solution, where the average decomposition cost is increased by 21.7\%. 
The reason is that some stitches which can be applied to solve conflicts are ignored by SDP-based algorithm, such that the average stitch number of SDP-based algorithm is reduced by 8.4\%.
Another approaching but not optimal algorithm is the dancing links based algorithm, where the decomposition cost is increased by 5.4\%.
The reason is that we only insert exactly one stitch candidate in each polygon feature for speedup while there are some features whose stitches are more than one,
which is ignored by our current dancing links implementation.

We also measure and compare the runtimes of different decomposition algorithms. 
The runtime result can be also found in \Cref{tab:total_result}, where the column ``time(s)" is the real time of decomposition in seconds instead of CPU time
considering that we use multiple threads in our experiments.
According to the results, backtracking-based algorithm faces a serious performance bottleneck when the input circuit becomes larger and
even fails to finish the decomposition procedure within 3600 seconds on \texttt{S35932} and \texttt{S38584} circuits.

Besides the worst backtracking algorithm, the other three algorithms also show an obvious differences in runtime, where the dancing links based algorithm 
outperforms all of the other algorithms in most cases and reduces the average runtime to 20.6\% compared with the ILP-based algorithm. 
SDP-based algorithm is also faster than ILP-based algorithm where the average runtime is reduced by 63.9\% but it is still almost twice as much as dancing links based algorithm.

\section{Conclusion and Future Work}
\label{sec:conclu}

$\mathsf{OpenMPL}$ is a general framework for multiple patterning layout decomposition problem and provides unified interfaces 
for layout decomposition algorithms and graph simplification speed-up techniques. 
Multi-threading, stitch insertion and shape-free feature are also supported in our framework. 
All of these features and some variables such as minimal distances are customizable and users can switch between these options freely.
This version of $\mathsf{OpenMPL}$ has implemented most state-of-the-art algorithms and the results demonstrate the effectiveness and the efficiency. 
However, there are still much room for $\mathsf{OpenMPL}$ to improve. 
In the future we plan to integrate post-refinement, optimize our stitch insertion phase
and develop acceleration techniques for the dancing links based algorithm.

\section{Acknowledgment}

This work is partially supported by
Cadence Design Systems, The Research Grants Council of Hong Kong SAR (No.~CUHK24209017), Taiwan MOST under Grant No.~106-2628-E-002-019-MY3, and US NSF under award No.~1718570.

\balance
{
    \bibliographystyle{IEEEtran}
    \bibliography{./ref/Top-sim,./ref/DFM,./ref/MPL,./ref/Software,./ref/liwei}
}

\end{document}